\documentclass[journal]{IEEEtran}
\usepackage{amsmath,graphicx,cite}
\usepackage{amsmath,amsthm,amssymb,mathtools}
\usepackage{upgreek}
\usepackage {tikz}
\usetikzlibrary {positioning,shapes,arrows,chains}
\usepackage{algorithmic}
\usepackage{algorithm}
\usepackage{scalerel}
\usepackage{pdfpages}

\tikzstyle{block} = [draw, fill=black!20, rectangle, 
minimum height=1em, minimum width=4em]
\tikzstyle{sum} = [draw, fill=black!20, circle, node distance=3 cm]
\tikzstyle{input} = [coordinate]
\tikzstyle{output} = [coordinate]
\tikzstyle{pinstyle} = [pin edge={to-,thin,black}]

\usepackage[utf8]{inputenc}
\usepackage[T1]{fontenc}

\usepackage{cite}

\ifCLASSINFOpdf
\else
\fi

\newcommand{\blankpage}{
\newpage
\thispagestyle{empty}
\mbox{}
\newpage
}

\ifCLASSOPTIONcompsoc
  \usepackage[caption=false,font=normalsize,labelfont=sf,textfont=sf]{subfig}
\else
  \usepackage[caption=false,font=footnotesize]{subfig}
\fi
\usepackage{url}

\begin{document}
%
\title{A Nonlinear Adaptive Filter Based on the Model of Simple Multilinear Functionals}
%
%
%

\author{\thanks{The authors are with the Department of Electronic Systems Engineering, Escola Politécnica,
University of São Paulo, São Paulo, Brazil (e-mails: felipe.chaud.pinheiro@usp.br;
cassio@lps.usp.br).}Felipe C. Pinheiro\thanks{The first author was supported by a scholarship from CNPq No.
132625/2015-6} and Cássio G. Lopes,~\IEEEmembership{Member,~IEEE}\thanks{The second author was supported by a grant from CNPq No.
311031/2013-7}}
\maketitle

\begin{abstract}
Nonlinear adaptive filtering allows for modeling of some additional aspects of a general system and usually relies on highly complex algorithms, such as those based on the Volterra series. Through the use of the Kronecker product and some basic facts of tensor algebra, we propose a simple model of nonlinearity, one that can be interpreted as a product of the outputs of K FIR linear filters, and compute its cost function together with its gradient, which allows for some analysis of the optimization problem. We use these results it in a stochastic gradient framework, from which we derive an LMS-like algorithm and investigate the problems of multi-modality in the mean-square error surface and the choice of adequate initial conditions. Its computational complexity is calculated. The new algorithm is tested in a system identification setup and is compared with other polynomial algorithms from the literature, presenting favorable convergence and/or computational complexity.\end{abstract}


%
\IEEEpeerreviewmaketitle

\section{Introduction}
%
%
%
%
\IEEEPARstart{T}{he} collective of nonlinear signal processing techniques come into play whenever nonlinear effects, such as harmonic distortion, saturation and general polynomial behavior, start to become noticeable enough to degrade the performance of conventional linear techniques. A common problem with nonlinear filters is the computational complexity they demand. Some of the most complex filters are based on the Volterra and Wiener models\cite{bk:ogunfunmi,poly,boyd}. It is common to restrict these models in order to reduce their complexity \cite{volten,batista1,batista2}. We follow this approach.

We start by proposing a polynomial technique that could be thought of as a subclass of the Volterra model. It is based on representing the input regressor as an iterated Kronecker product, an idea that has already been used for the estimation of higher order statistics of the regressor \cite{high}---a related problem---and other Volterra approaches \cite{volten,batista1,batista2}, but we also embed this Kronecker structure within the filter itself. This formalism allows us to promptly derive a cost function in explicit form. We also compute its gradient and use it in the derivation of a nonlinear, low-complexity, LMS-like algorithm with good mean-square performance.

\section{Simple Multilinear Model}

The general $K$-th order homogeneous Volterra kernel is a function $h(i_1, \dotsc, i_K)$ used to produce the input/output relationship in \eqref{eq:hom}. There, $u(i)$ is the input signal and $M$ is the length of the filter.
\begin{equation}\label{eq:hom}
y(i) = \sum_{0 \le i_1,\dotsc,i_K < M} h(i_1, \dotsc, i_K) u(i - i_1) \dotsb u(i - i_K)
\end{equation}

Evaluating \eqref{eq:hom} is a highly complex task. One would take $O(M^K)$ operations per iteration. This can be slightly reduced should the symmetry in the kernel be explored, but this approach could at most decrease the number of operations to something proportional to $M-K+1\choose K$, which still increases rapidly. It is necessary to suppose some extra structure in the kernel itself if one wants a significant reduction in complexity.

The structure we propose is kernel separability. Explicitly, we suppose that there exists functions $h_s(i_s)$, $1\le s \le K$, such that 
\begin{equation}\label{eq:sep}
h(i_1, \dotsc, i_K) = h_1(i_1) h_2(i_2) \dotsm h_K(i_K).
\end{equation}

When we use \eqref{eq:sep} in \eqref{eq:hom}, we get
\begin{align}\label{eq:simple_hom}
y(i) &= \sum_{0 \le i_1,\dotsc,i_K < M} h_1(i_1) \dotsm h_K(i_K) u(i - i_1) \dotsm u(i - i_K)\nonumber\\
&= \sum_{i_1=0}^{M-1} h_{1}(i_1) u(i - i_1) \dotsb \sum_{i_K=0}^{M-1} h_{K}(i_K) u(i - i_K).
\end{align}

Each factor $y_{os}(i) \triangleq \sum_{i_s=0}^M h_{s}(i_j) u(i - i_s)$ could be seen as the output of a linear FIR filter. As such, we can represent them in vector form. We collect the input signals in a $1 \times M$ row vector\footnote{This notation is due to \cite{bk:sayed}. Scalars are represented as $x(i)$, vectors as $x_i$, matrices and constants as capital letters (either is clear from context) and a boldface font for random quantities.}  $u_i = [u(i)\,u(i-1) \cdots u(i - M + 1)]$ and a set of $K$ vectors of size $M\times 1$ represented by $\{w_1, \dotsc, w_K\}$. We also say that, for each $s$, we have 
\begin{equation}
w_s = [h_s(0)\, h_s(1)\cdots h_s(M-1)]^T.
\end{equation}
Now, we have $y_{os}(i) = u_i w_s$ for every $s$, so that 
\begin{equation}\label{eq:mult}
y(i) = (u_i w_{1}) (u_i w_{2}) \dotsb (u_i w_{K}).
\end{equation}

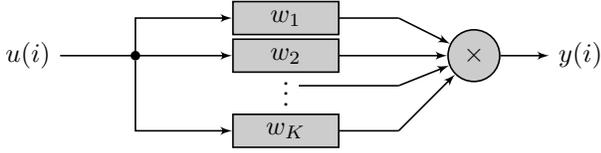
\begin{figure}[h]
\pgfdeclarelayer{background layer}
\pgfdeclarelayer{foreground layer}
\pgfsetlayers{background layer,main,foreground layer}
\centering
\begin{tikzpicture}[auto, node distance=0.5cm,>=latex', semithick]

\node [input, label = left:$u(i)$] (input) {};
\node [circle, fill = black, draw = black, minimum size=3pt, inner sep=0pt, node distance = 1.0 cm, right of = input] (inputbranch) {};
\node [block, right of=input, node distance = 3.0 cm] (afilter) {$w_2$};
\node [sum, right of = afilter, node distance = 2.5 cm] (diff) {$\times$};

\node [block, above of=afilter] (system) {$w_1$};
\node [below of = afilter, node distance = 0.4 cm] (dot) {$\vdots$};
\node [block, below of=dot, node distance = 0.6 cm] (wk) {$w_K$};

\node [coordinate, right of = system, node distance = 1.5 cm] (o1) {$\vdots$};
\node [coordinate, right of = wk, node distance = 1.5 cm] (ok) {$\vdots$};
\node [coordinate, right of = dot, node distance = 1.5 cm] (od) {$\vdots$};

\node [output, right of=diff,node distance = 1.0 cm, label = right:$y(i)$] (out) {};

\draw [->] (input) -- node {} (afilter);
\draw [->] (afilter) -- node {} (diff);
\draw [->] (inputbranch) |- node {} (system);
\draw [->] (inputbranch) |- node {} (wk);
\draw [-] (system) -- node {} (o1);
\draw [->] (o1) -- node {} (diff);
\draw [-] (wk) -- node {} (ok);
\draw [->] (ok) -- node {} (diff);
\draw [-] (dot) -- node {} (od);
\draw [->] (od) -- node {} (diff);
\draw [->] (diff) -- node {} (out);
\end{tikzpicture}
\caption{Block diagram of the Simple Multilinear Model.}
\label{fig:sml}
\end{figure}

This can be represented by Fig. \ref{fig:sml}. Granted, this implies a loss of generality from the original Volterra model. Not every kernel is separable. But this simplification allows us to compute the output of the system in $O(KM)$ operations per iteration---a exponential reduction in complexity. Moreover, this separable kernel allows a convenient formalism for the derivation of the cost function that also gives us algebraic insight on the model.

\subsection{Kronecker Representation}

Using the identity $(A \otimes B) (C \otimes D) = (AC) \otimes (BD)$ for the Kronocker product \cite{kron}, it follows that \eqref{eq:mult} is equivalent to \eqref{eq:tensor}, because the Kronecker product of scalars is their ordinary multiplication.
\begin{align}\label{eq:tensor}
y(i) &= ( u_i w_1 )\dotsm ( u_i w_K )= ( u_i w_1) \otimes \dotsb \otimes ( u_i w_K ) \nonumber\\
&= \underbrace{(u_i \otimes \dotsb \otimes u_i)}_{K\text{ times}} (w_1 \otimes \dotsb \otimes w_K) 
\end{align}

The Kronecker product is analogous to the tensor product \cite{bk:roman}, therefore we can think of the two factors $(w_1 \otimes \dotsb \otimes w_K)$ and $(u_i \otimes \dotsb \otimes u_i)$ in \eqref{eq:tensor} as tensors. As such, to better manipulate and have access to their elements, one can even index them as
\begin{equation}\label{eq:uind}
(u_i \otimes \dotsb \otimes u_i)_{j_1, \dotsc,j_K} = u(i - j_1 + 1)\dotsm u(i - j_K + 1)
\end{equation} 
and
\begin{equation}\label{eq:wind}
(w_1 \otimes \dotsb \otimes w_K)^{i_1, \dotsc,i_K} = h_1(i_1)\dotsm h_K(i_K).
\end{equation} 

This should be so that, if we sum the product of \eqref{eq:uind} and \eqref{eq:wind} over the indexes $i_s = j_s$ for every $s$, we get the output of the system: \footnote{This fact is used in Appendix A.}
\begin{equation}
\sum_{i_1,\dotsc,i_K} (u_i \otimes \dotsb \otimes u_i)_{i_1, \dotsc,i_K}(w_1 \otimes \dotsb \otimes w_K)^{i_1, \dotsc,i_K} = y(i).
\end{equation}

Actually, these objects are a very specific kind of tensor: they have rank one and are also called \emph{decomposable} or \emph{simple}. One could think of substituting $w_1 \otimes \dotsb \otimes w_K$ for a general tensor and produce much more diverse nonlinearities. However, such procedure would lead back to the full Volterra model. This fact sheds algebraical reasoning on what can be represented by the a product of FIR systems.

Additionally, one can think of the ``$w$'' tensor as a $K$-linear form\cite{greub}---or functional---acting on $K$ copies of the vector $u_i$. In this case, they are rank one multilinear functionals, also called \emph{simple multilinear functionals}, from which we take the name of our model: the simple multilinear model (SML).

\section{The Mean Square Error over the SML}

As it is usual in most adaptive schemes, we pose a problem related to optimization---in particular, the minimization of a certain metric, the most common being the mean-square error.

%
%
%
%

Given a desired random signal $\mathbf{d}$, a random $1\times M$ vector $\mathbf{u}$ and a set of $M\times 1$ column vectors $\{w_1, \dotsc, w_K\}$, the output estimation error is defined as $\mathbf{e} \triangleq \mathbf{d} - \mathbf{u}^{\otimes K} w$. Under our notation, $w \triangleq w_1 \otimes \dotsb \otimes w_K$ ($KM\times 1$), and $\mathbf{u}^{\otimes K} \triangleq \mathbf{u} \otimes \dotsb \otimes \mathbf{u}$ ($1\times KM$), the Kronecker product of $\mathbf{u}$ with itself $K$ times. The mean-square error (MSE) is defined as
\begin{align}\label{eq:mse_eval}
\text{MSE} &\triangleq \mathbb{E}|\mathbf{e}|^2 = \mathbb{E}\left[[\mathbf{d} - \mathbf{u}^{\otimes K}w]^* [\mathbf{d} - \mathbf{u}^{\otimes K}w]\right] \nonumber\\
&=  \mathbb{E}|\mathbf{d}|^2 - w^*\mathbb{E}[\mathbf{d} \mathbf{u}^{\otimes K*}] - \mathbb{E}[\mathbf{d}^* \mathbf{u}^{\otimes K}] w \\
&\quad + w^* \mathbb{E}[\mathbf{u}^{\otimes K*} \mathbf{u}^{\otimes K}] w.\nonumber
\end{align}

Under the hypothesis of stationarity on both $\mathbf{u}$ and $\mathbf{d}$, we can define the following constants:
\begin{gather}
R_{u^K} = \mathbb{E}[\mathbf{u}^{\otimes K*} \mathbf{u}^{\otimes K}],\, 
R_{u^K d} = \mathbb{E}[\mathbf{u}^{\otimes K} \mathbf{d}^*]  = R_{du^K}^*\label{eq:rdu}\\
R_d = \mathbb{E}|\mathbf{d}|^2.
\end{gather}

Then, the mean square error takes the familiar form in \eqref{eq:mse}.
\begin{align}\label{eq:mse}
\text{MSE}(w_1,& \dotsc, w_K) = \nonumber\\ 
                  &R_d - w^* R_{u^K d}^* - R_{u^K d} w + w^* R_{u^K} w.
\end{align}

Although the form is familiar, our notation embeds some aspects of this function. For example, it does not describe a quadratic surface. Instead, it has degree $2K$, when we look at the individual vectors $w_1, \dotsc, w_K$. 

Note that \eqref{eq:mse} could be seen as a function of a vector $w^C$ built from the stacking of the individual $w_1, \dotsc, w_K$ vectors. We now introduce $\nabla_{w_s}$, the gradient with respect to the coordinates of the vector $w_s$. The complete gradient vector, $\nabla\text{MSE}$, over $w^C$, would be the stacking of these gradients. It is possible to show that (see Appendix A)
\begin{align}\label{eq:grad}
\nabla_{w_s} \text{MSE} = [-R_{u^K d} + w^* R_{u^K}] (w_1 \otimes \dotsb \otimes \widehat{w_s }\otimes \dotsb \otimes w_K),
\end{align}
where $\widehat{w_s}$ implies that $w_s$ has been substituted for the identity matrix $I_M$ of order $M$ in the product. 

In addition, the critical MSE is reached when the gradient is $0$. This implies the trivial solution $w_p = w_q = 0$, for some $p\ne q$---which makes $w_1 \otimes \dotsb \otimes \widehat{w_s }\otimes \dotsb \otimes w_K = 0$ for each $s$---or the equation $R_{u^K} w_o = R_{du^K} $. Usually we are not interested in the first, therefore we will focus on the second. This equation is reminiscent of the normal equations in linear estimation. As such, we can reorganize it in the form
\begin{gather}
\mathbb{E}[\mathbf{u}^{\otimes K} \mathbf{u}^{\otimes K*}] w_o = \mathbb{E}[\mathbf{u}^{\otimes K*} \mathbf{d}] \nonumber\\
\mathbb{E}[\mathbf{u}^{\otimes K*} (\mathbf{d} - \mathbf{u}^{\otimes K}w_o)] = 0\nonumber\\
\mathbb{E}[\mathbf{u}^{\otimes K*}\mathbf{e}_o] = 0,
\end{gather}
that is, a form of the orthogonality principle, with the optimal output error $\mathbf{e}_o$ being orthogonal to $\mathbf{u}^{\otimes K}$. One can interpret this as $\mathbf{e}_o$ being uncorrelated to all of the ``degree $K$'' products of the input.

Due to the structure of $R_{u^K}$ (repeated rows), it will always be singular for $K>1$, but under certain conditions we can guarantee the existence of the non trivial solution. Assume for $\mathbf{d}$ the model in \eqref{eq:dmodel},
\begin{equation}\label{eq:dmodel}
\mathbf{d} = \mathbf{u}^{\otimes K}h_o + \mathbf{n},
\end{equation}
 where $\mathbf{n}$ is some zero-mean noise uncorrelated with $\mathbf{u}^{\otimes K}$ (serving as a model for $\mathbf{e}_o$), and $h_o = h_{o1} \otimes \dotsb \otimes h_{oK}$, for some set of vectors $\{h_{o1},\dotsc,h_{oK}\}$. Then, one can verify that $h_o$ is a solution to the equation $R_{u^K} w_o = R_{du^K}$ in $w_o$:
\begin{align}
R_{du^K} &=  E[\mathbf{u}^{\otimes K*}\mathbf{d}] =  E[\mathbf{u}^{\otimes K*} \mathbf{u}^{\otimes K}]h_o +  E[ \mathbf{u}^{\otimes K*}\mathbf{n}] \nonumber\\
&=  R_{u^K}h_o .
\end{align}

\section{An LMS Inspired Algorithm}
We take the total gradient $\nabla\text{MSE}$ over $w^C$, as previously defined. Then, we use it to form the equation of the steepest descent.
\begin{equation}\label{eq:step}
w^C[i] = w^C[i-1] - \mu [\nabla\text{MSE}]^*.
\end{equation}

As it is usually done for stochastic algorithms, we use the realizations $u_i$ and $d(i)$ to estimate \eqref{eq:rdu}.
\begin{equation}
\widetilde{R}_{u^K} = u_i^{\otimes K*} u_i^{\otimes K},\quad
\widetilde{R}_{u^K d} = u_i^{\otimes K} d(i)^*
\end{equation}

and compute the gradient accondingly from \eqref{eq:grad}. We define $y_{s}(i) \triangleq (u_i w_1) \dotsm \widehat{(u_i w_s)}\dotsm (u_i w_K)$, where $\widehat{(u_i w_s)}$ denotes the absence of the factor $(u_i w_s)$ in the product. Then
\begin{align}\label{eq:estgrad}
\widetilde{\nabla}_{w_s} \text{MSE} &= [-d(i)^* u_i^{\otimes K} + w^* u_i^{\otimes K*}u_i^{\otimes K}] \nonumber\\
&\quad\cdot (w_{1} \otimes \dotsb \otimes\widehat{ w_{s} }\otimes \dotsb \otimes w_{K})\nonumber\\
&= - [d(i) - u_i^{\otimes K} w]^*(u_i \otimes \dotsb \otimes u_i) \nonumber\\
&\quad\cdot (w_1 \otimes \dotsb \otimes I_M \otimes \dotsb \otimes w_K) \nonumber\\
&= - e(i)^* (u_i w_1) \dotsb (u_i I_M) \dotsb (u_i w_K)  \nonumber\\
&= - (u_i w_1) \widehat{\dotsb (u_i w_s) \dotsb} (u_i w_K) u_i e(i)^* \nonumber\\
&= - y_{s}(i) e(i)^* u_i.
\end{align}

It follows, then, by using \eqref{eq:estgrad} in \eqref{eq:step}, the update law in \eqref{eq:update}, that should be performed for every $s, 1\le s\le K$.
\begin{align}\label{eq:update}
w_s{[i]} = w_s{[i-1]}+ \mu e(i) y_{s}(i)^*  u_i^*
 \end{align}


The algorithm still needs to initialize its variables---and this is an important point in the implementation. As it can be seen from the MSE surface, $w_1 = \dotsb = w_K = 0$ is a critical point. Correspondingly, if every $w_j$ is initialized at $0$, \eqref{eq:update} shows that they will always be $0$. 
In spite of these problems, we have empirically found one satisfactory initial condition for use in high numeric precision environments. And that is: using $w_1[0] = [1\,0\,\cdots\, 0\,0]^T$ , $w_2[0] = [2^{-1}\,0\,\cdots\, 0\,0]^T$, or generally $w_j[0] = [2^{1 - j}\,0\,\cdots\, 0\,1]^T$ until $j = K-1$ and then $w_K[0] = 0$. This is all summarized in the Algorithm Table \ref{tb:algo}.


\begin{algorithm}[t]                  
\caption{LMS-like algorithm}          
\label{tb:algo}                           
\footnotesize
\begin{algorithmic}                    
\STATE \textbf{Initialization} \\
\FOR{$j = 1$ to $K-1$}
	\STATE $w_j[0] = [2^{1-j}\,0\,\cdots\, 0\,0]^T$ \\
\ENDFOR
\STATE $w_K[0] = [0\,0\,\cdots\, 0\,0]^T$ \\
\STATE \textbf{Iteration}
\FOR{$i = 1$ to END}
	\FOR{$s = 1$ to $K$}
		\STATE {Compute $y_{os}(i) = u_i w_s[i-1]$}
	\ENDFOR
	\FOR{$s = 1$ to $K$}
		\STATE {Compute $y_{s}(i) = y_{o1}(i) \widehat{\dotsb y_{os}(i) \dotsb} y_{oK}(i)$}
	\ENDFOR
	\STATE {Compute $y(i) = y_{K}(i) y_{oK}(i)$}
	\STATE {Compute $e(i) = d(i) - y(i)$}
	\STATE {Compute $f_i = \mu e(i) u_i^*$}
	\FOR{$s = 1$ to $K$}
		\STATE {Compute $w_s{[i]} = w_s{[i-1]} + f_i y_{s}(i)^* $}
	\ENDFOR
\ENDFOR
\end{algorithmic}
\end{algorithm}

To calculate the computational complexity for algorithm \eqref{eq:update}, focus on the multiplications present in Algorithm Table 1. The total number of them, on a given iteration, are, for real data and $K \ge 2$, given by $M K + K^2 - K + 2 M + 2$. In other words, this an $O(M K  + K^2)$ algorithm.

\section{Simulations}
\begin{figure}[!t]
\centering
\subfloat[Case I: Algorithms identifying an order 2 SML plant.]{
\includegraphics[clip = true, trim = 200 315 200 315, width=1.40in]{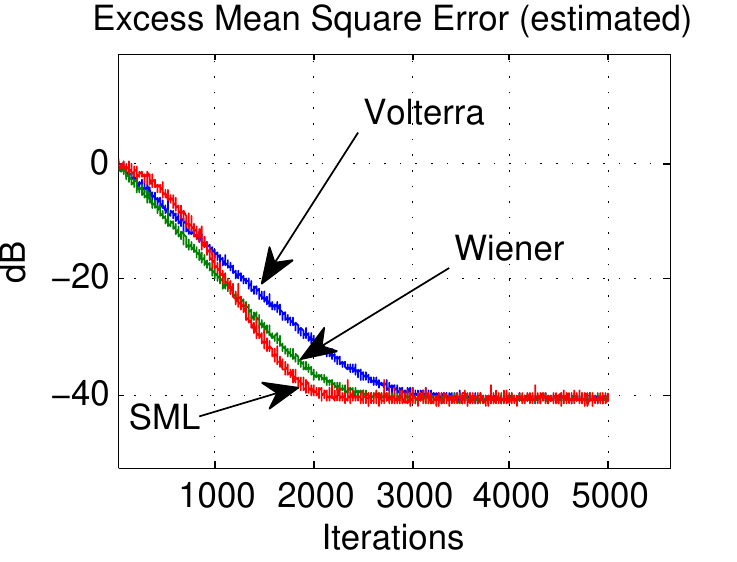}%
\label{ploti}
}\hspace{2mm}
\subfloat[Case II: Same as Case I; higher SNR.]{
\includegraphics[clip = true, trim = 200 315 200 315, width=1.40in]{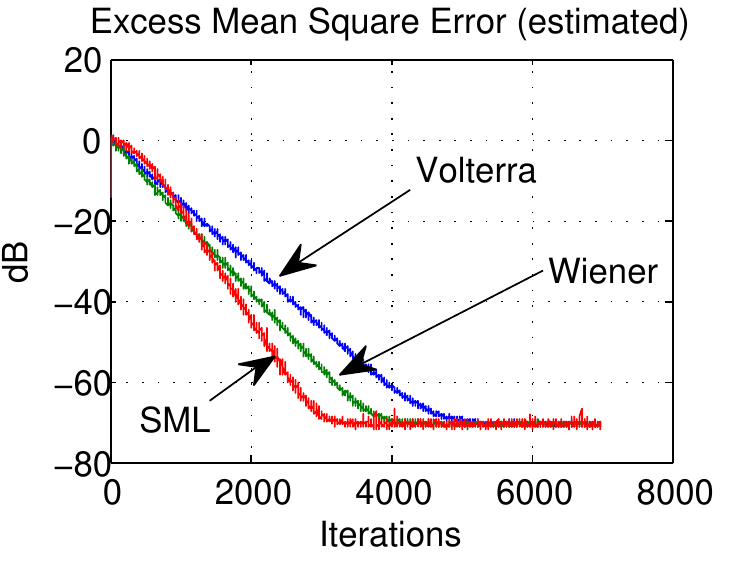}%
\label{plotii}
}

\subfloat[Case III: Algorithms identifying a different order 2 SML plant.]{
\includegraphics[clip = true, trim = 200 315 200 315, width=1.40in]{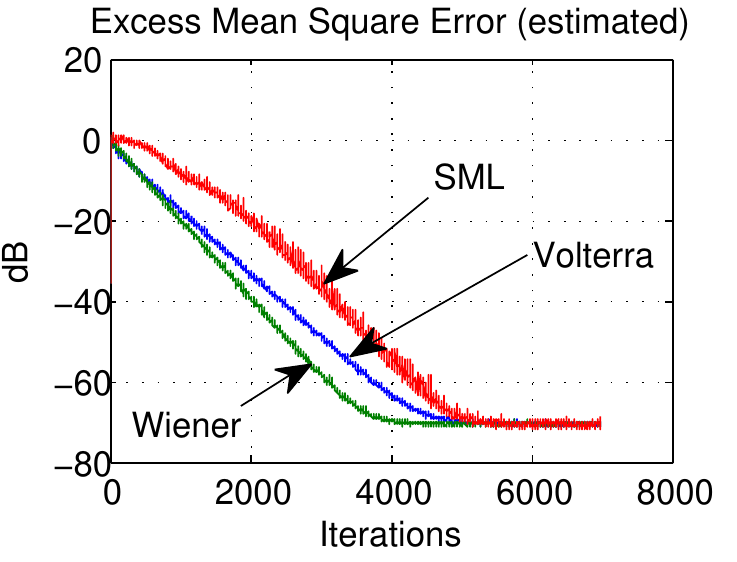}%
\label{plotiii}
}\hspace{2mm}
\subfloat[Case IV:  Algorithms identifying an order 3 SML plant.]{
\includegraphics[clip = true, trim = 200 315 200 315, width=1.40in]{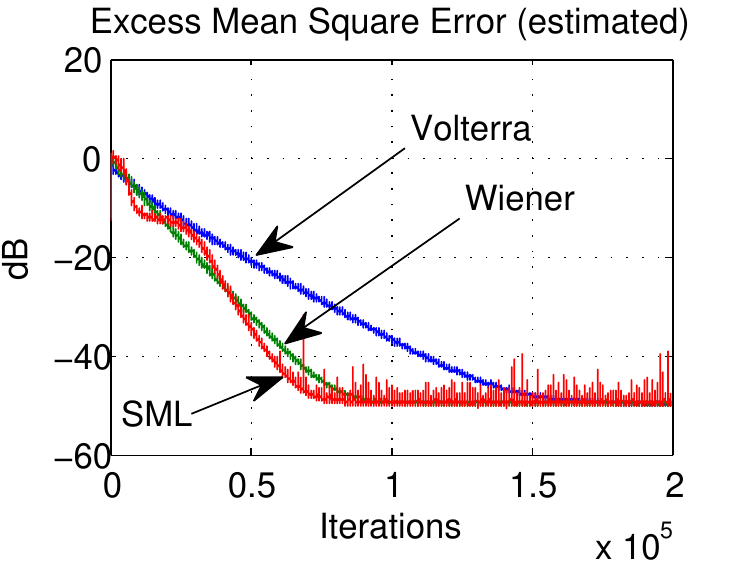}%
\label{plotiv}
}

\subfloat[Case V: Various filters identifying a smooth plant.]{
\includegraphics[clip = true, trim = 200 315 200 325, width=1.50in]{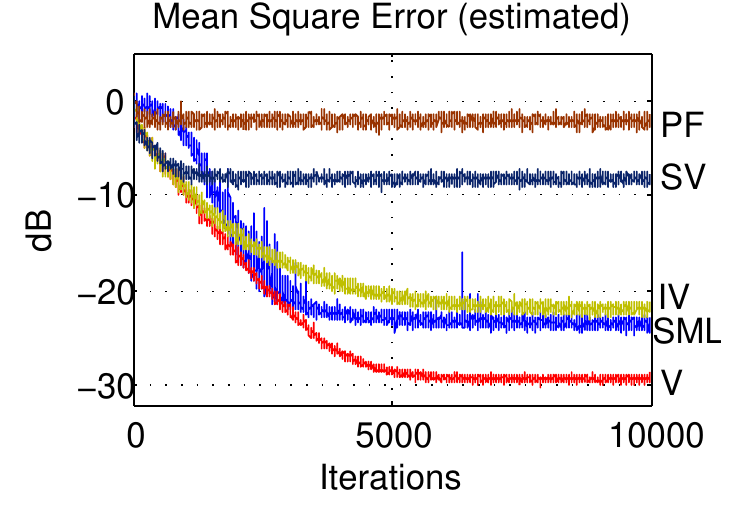}%
\label{plotv}
}\hspace{2mm}
\subfloat[Case VI: Parallel cascade and SML.]{
\includegraphics[clip = true, trim = 200 315 200 325, width=1.50in]{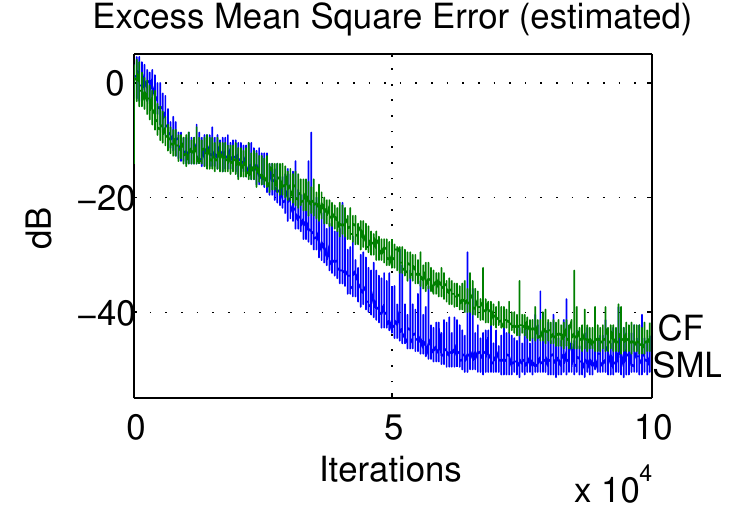}%
\label{plotvi}
}
\caption{Curves of the MSE and ESME showing the performance of the SML algorithm against other polynomial algorithms.}
\label{fig:plot}
\end{figure}

A set of two simulations, Cases I and II, on the problem of system identification, with $K = 2$ and $M = 10$, were run. The signal $d(i)$ was an SML plant plus additive zero-mean Gaussian noise with variance $\sigma_n^2$ (see \eqref{eq:dmodel}). The input signal was drawn from a zero-mean Gaussian distribuition with unitary variance, collected in the vector $u_i$ with a delay-line structure. 
In Case I, we chose $\sigma_n^ 2 = 10^{-3}$ and in Case II, $\sigma_n^2 = 10^{-6}$. In each case, the algorithm was run through 7000 iterations and the curves were averaged through 1000 realizations. The resulting Excess Mean Square Error (EMSE) 
curves 
(calculated as $\mathbb{E}|\mathbf{u}^{\otimes 2}(h_1\otimes h_2 - \mathbf{w}[i-1])|^ 2$) 
 are present in Figs. \ref{ploti} and \ref{plotii}. Plotted together on the graph are the curves for the Volterra-LMS and the Wiener-LMS, designed for the Gaussian vector $\mathbf{u}$, as described in \cite{bk:ogunfunmi}. In those algorithms, we chose\footnote{Stability bounds require a more sophisticated analysis.} $\mu$
so to get an approximately equal steady-state EMSE.

%

%

%
%
%


For Case III we have used a different SML plant. We had $\sigma_n^2 = 10^{-6}$. 
$\mu$ was again chosen to equalize the steady-state EMSE. The rest of the parameters remained the same.



The algorithm was also made to run with $K = 3$, which makes Case IV. We had
$M = 10$ and 200,000 iterations, averaged through 10,000 realizations. We compare the results with the Volterra-LMS
and the Wiener-LMS
with parameters that equalize the steady-state EMSE. The resulting curves are presented in Fig. \ref{plotiv}. The variance of the additive noise was chosen as $\sigma_n^2 = 10^{-3}$.


Case V compares the SML-LMS with some relevant LMS-like polynomial algorithms from the literature. Fig. \ref{plotv} shows the Power Filter (PF) \cite{pfilter}, Simplified Volterra (SV) \cite{svolt,svolt2}, Sparse Interpolated Volterra (IV) \cite{batista1,batista2}, the regular Volterra filter and the SML. The task was to identify a smooth, non-SML, order 2 Volterra plant with $M = 21$. This was averaged through 1000 realizations.

Case VI shows a simulation against another filter, the Parallel Cascade Filter (CF) \cite{cascade}---single branch and LMS version. This algorithm is similar to the SML for $K \le 2$, so we simulated them in $K=3$, where they are clearly dissimilar. We identify the same plant as Case IV, with $M = 10$, through 1000 realizations. The results are on Fig. \ref{plotvi}. 

For both Cases V and VI the additive noise variance was $10^{-3}$. The parameters were adjusted to provide approximately the same convergence rate.

A description of all the plants is available with the on-line materials in the file plants.pdf.

\section{Discussion}

Cases I and II clearly show the system converging to an EMSE of approximately $-40$ dB and $-70$ dB, respectively. On both Figs. \ref{ploti} and \ref{plotii}, we can see the SML-LMS running slower through the first 500 iterations, then it speeds up and catches up with---and eventually overtake---the others. Both the Volterra and Wiener models assume $O\left(M+ K - 1 \choose K\right)$ operations per iteration, while the SML only $O(K M + K^2)$. Although it is natural that the SML algorithm is better at identifying SML plants, it does it faster and with far fewer arithmetic operations.

Note that the SML algorithm presents richer dynamics compared to the other algorithms---something that will be studied in following publications. Case III, in Fig. \ref{plotiii}, for example, shows that
 its behavior does not depend only on statistics of the regressor $u_i$---in fact, it is also influenced by 
 the plant to be identified. A source of these effects can be hypothesised as due to the $y_s(i)$ factors in \eqref{eq:update}. 

Case IV, in Fig. \ref{plotiv}, shows that the algorithm also converges for $K = 3$ and that it has the aforementioned initial irregular convergence rate, here appearing in a more evident way.

Case V shows that the algorithm is competitive, even in a non-SML plant---as long as this plant is correlated enough. The PF used 21 coefficients, the SV used 66, the IV used 60, Volterra used 231 and SML 42. This is another result that shows the computational simplicity of the SML.

Case VI shows two similar filters, in the sense that they are formed through the product of others. SML is a product of three linear filters, with a total of 30 coefficients, and CF is a product of a linear filter with a second order Volterra filter, with 66 coefficients. In the task of identifying an SML plant, they show similar behavior, with the SML being slightly better---at half the complexity. 


\section{Conclusion}
 Through the use of a Kronecker representation, we have developed the cost function for the SML model. From this function its gradient was derived motivating a new LMS-like nonlinear adaptive algorithm. It has very low computational complexity---an exponential reduction when compared to te Volterra series---and a competitive mean-square performance in a variety of cases, but also some other unusual behavior.

Finally, the SML algorithm, just like the Parallel Cascade Filter \cite{cascade}, allows the extension for models of greater rank. This path will be pursued in future publications.


%

\appendices
\section{Proof of the Gradient Formula}
Through \eqref{eq:uind} and \eqref{eq:wind}, one can interpret \eqref{eq:grad} as a tensor equation, in the sense that we can index the matrices as
\[
(R_{u^ K})^ {i_1,\dotsc,i_K}_{j_1,\dotsc,j_K}=\mathbb{E}(\mathbf{u}^{\otimes K *}\mathbf{u}^{\otimes K})^ {i_1,\dotsc,i_K}_{j_1,\dotsc,j_K}
\]
\[
\text{and}\quad(R_{u^Kd})_{j_1,\dotsc,j_K} = \mathbb{E}(\mathbf{u}^{\otimes K }\mathbf{d})_{j_1,\dotsc,j_K}
\]
so to write
\[
R_{u^Kd}w = \sum_{j_1,\dotsc,j_K}(R_{u^Kd})_{j_1,\dotsc,j_K}\prod_\ell (w_\ell)^{j_\ell},
\]
where $(w_\ell)^{j_\ell}$ is the $j_\ell$-th coordinate of $w_\ell$, and
\[
w^*R_{u^K}w = \sum_{\substack{i_1,\dotsc,i_K\\j_1,\dotsc,j_K}}\prod_{p} (w_p)^{i_{p}*}(R_{u^K})^{i_1,\dotsc,i_K}_{j_1,\dotsc,j_K}\prod_{\ell} (w_\ell)^{j_{\ell}}.
\]

The other terms from \eqref{eq:grad} involve only the conjugates of the entries of $w$, so they became to zero\cite{bk:sayed}.

The gradient over the vector $w_{s}$ is given by the derivatives over each of its components:
\[
\frac{\partial (R_{u^Kd})w}{\partial(w_s)^{j_q}} = \sum_{j_1,\dotsc,j_K}(R_{u^Kd})_{j_1,\dotsc,j_K}\prod_{\ell\ne s} (w_\ell)^{j_\ell}\delta^{j_s}_{j_q}.
\] 

$\delta^i_j$ is the Kronecker delta and also the representation of the identity matrix. In the above expression, the delta---the identity---takes the place of the factor $(w_s)^{j_s}$. This observation leads directly to
\[
\nabla_{w_s} (R_{u^Kd}w) =  R_{u^Kd} (w_1 \otimes \dotsb \otimes\widehat{ w_s}\otimes \dotsb \otimes w_K).
\] 

For $w^*R_{u^K}w$, we have, remembering we do not derivate the conjugates \cite{bk:sayed},
\[
\frac{\partial (w^*R_{u^K}w)}{\partial(w_s)^{j_q}} = \sum_{\substack{i_1,\dotsc,i_K\\j_1,\dotsc,j_K}}\prod_{p} (w_p^*)_{i_{p}}(R_{u^K})^{i_1,\dotsc,i_K}_{j_1,\dotsc,j_K}\prod_{\ell\ne s} (w_\ell)^{j_{\ell}}\delta^{j_s}_{j_q}.
\]

Under the same argument, 
\[
\nabla_{w_s} (wR_{u^K}w^*) = w^* R_{u^K} (w_1 \otimes \dotsb \otimes \widehat{w_s}\otimes  \dotsb \otimes w_K).
\]

Therefore, we combine those two terms to get
\begin{align*}
\nabla_{w_s}\text{MSE} &= \nabla_{w_s} (R_d - w^* R_{u^K d}^*  - R_{u^K d} w + w^* R_{u^K} w)\nonumber\\
&= -\nabla_{w_s} (R_{u^K d} w) + \nabla_{w_s} (w^* R_{u^K} w)\nonumber\\
&= [-R_{u^K d} + w^* R_{u^K}](w_1 \otimes \dotsb \otimes \widehat{w_s}\otimes \dotsb \otimes w_K).
\end{align*}

\ifCLASSOPTIONcaptionsoff
  \newpage
\fi



\bibliographystyle{IEEEtran}
\bibliography{IEEEfull,refs}

\begin{thebibliography}{10}
\providecommand{\url}[1]{#1}
\csname url@samestyle\endcsname
\providecommand{\newblock}{\relax}
\providecommand{\bibinfo}[2]{#2}
\providecommand{\BIBentrySTDinterwordspacing}{\spaceskip=0pt\relax}
\providecommand{\BIBentryALTinterwordstretchfactor}{4}
\providecommand{\BIBentryALTinterwordspacing}{\spaceskip=\fontdimen2\font plus
\BIBentryALTinterwordstretchfactor\fontdimen3\font minus
  \fontdimen4\font\relax}
\providecommand{\BIBforeignlanguage}[2]{{%
\expandafter\ifx\csname l@#1\endcsname\relax
\typeout{** WARNING: IEEEtran.bst: No hyphenation pattern has been}%
\typeout{** loaded for the language `#1'. Using the pattern for}%
\typeout{** the default language instead.}%
\else
\language=\csname l@#1\endcsname
\fi
#2}}
\providecommand{\BIBdecl}{\relax}
\BIBdecl

\bibitem{bk:ogunfunmi}
T.~Ogunfunmi, \emph{Adaptive Nonlinear System Indentification: The Volterra and
  Wiener Model Approaches}.\hskip 1em plus 0.5em minus 0.4em\relax Secaucus,
  NJ, USA: Springer-Verlag New York, Inc., 2006.

\bibitem{poly}
V.~Mathews, ``Adaptive polynomial filters,'' \emph{Signal Processing Magazine,
  IEEE}, vol.~8, no.~3, pp. 10--26, July 1991.

\bibitem{boyd}
S.~Boyd, L.~O. Chua, and C.~A. Desoer, ``Adaptive polynomial filters,''
  \emph{IMA Journal of Mathematical Control and Information, Oxford University
  Press}, vol.~1, no.~3, pp. 243--282, 1984.

\bibitem{volten}
R.~Nowak and B.~Van~Veen, ``Tensor product basis approximations for volterra
  filters,'' \emph{Signal Processing, IEEE Transactions on}, vol.~44, no.~1,
  pp. 36--50, Jan 1996.

\bibitem{batista1}
E.~Batista, O.~J. Tobias, and R.~Seara, ``A fully lms adaptive interpolated
  volterra structure,'' in \emph{Acoustics, Speech and Signal Processing, 2008.
  ICASSP 2008. IEEE International Conference on}, March 2008, pp. 3613--3616.

\bibitem{batista2}
E.~Batista, O.~Tobias, and R.~Seara, ``A sparse-interpolated scheme for
  implementing adaptive volterra filters,'' \emph{Signal Processing, IEEE
  Transactions on}, vol.~58, no.~4, pp. 2022--2035, April 2010.

\bibitem{high}
T.~Andre, R.~Nowak, and B.~Van~Veen, ``Low rank estimation of higher order
  statistics,'' in \emph{Acoustics, Speech, and Signal Processing, 1996.
  ICASSP-96. Conference Proceedings., 1996 IEEE International Conference on},
  vol.~5, May 1996, pp. 3026--308a vol. 5.

\bibitem{bk:sayed}
A.~H. Sayed, \emph{Adaptive Filters}.\hskip 1em plus 0.5em minus 0.4em\relax
  Wiley-IEEE Press, 2008.

\bibitem{kron}
J.~Brewer, ``Kronecker products and matrix calculus in system theory,''
  \emph{Circuits and Systems, IEEE Transactions on}, vol.~25, no.~9, pp.
  772--781, Sep 1978.

\bibitem{bk:roman}
S.~Roman, \emph{Advanced Linear Algebra}.\hskip 1em plus 0.5em minus
  0.4em\relax Springer, 2007.

\bibitem{greub}
W.~Greub, \emph{Multilinear Algebra}, ser. Universitext.\hskip 1em plus 0.5em
  minus 0.4em\relax Springer New York, 2012.

\bibitem{pfilter}
F.~Kuech, A.~Mitnacht, and W.~Kellermann, ``Nonlinear acoustic echo
  cancellation using adaptive orthogonalized power filters,'' in
  \emph{Acoustics, Speech, and Signal Processing, 2005. Proceedings. (ICASSP
  '05). IEEE International Conference on}, vol.~3, March 2005, pp.
  iii/105--iii/108 Vol. 3.

\bibitem{svolt}
\BIBentryALTinterwordspacing
A.~Fermo, A.~Carini, and G.~L. Sicuranza, ``Low-complexity nonlinear adaptive
  filters for acoustic echo cancellation in gsm handset receivers,''
  \emph{European Transactions on Telecommunications}, vol.~14, no.~2, pp.
  161--169, 2003. [Online]. Available: \url{http://dx.doi.org/10.1002/ett.908}
\BIBentrySTDinterwordspacing

\bibitem{svolt2}
------, ``Simplified volterra filters for acoustic echo cancellation in gsm
  receivers,'' in \emph{Signal Processing Conference, 2000 10th European}, Sept
  2000, pp. 1--4.

\bibitem{cascade}
T.~M. Panicker, V.~J. Mathews, and G.~L. Sicuranza, ``Adaptive parallel-cascade
  truncated volterra filters,'' \emph{IEEE Transactions on Signal Processing},
  vol.~46, no.~10, pp. 2664--2673, Oct 1998.

\end{thebibliography}
\blankpage

\includepdf[pages=-]{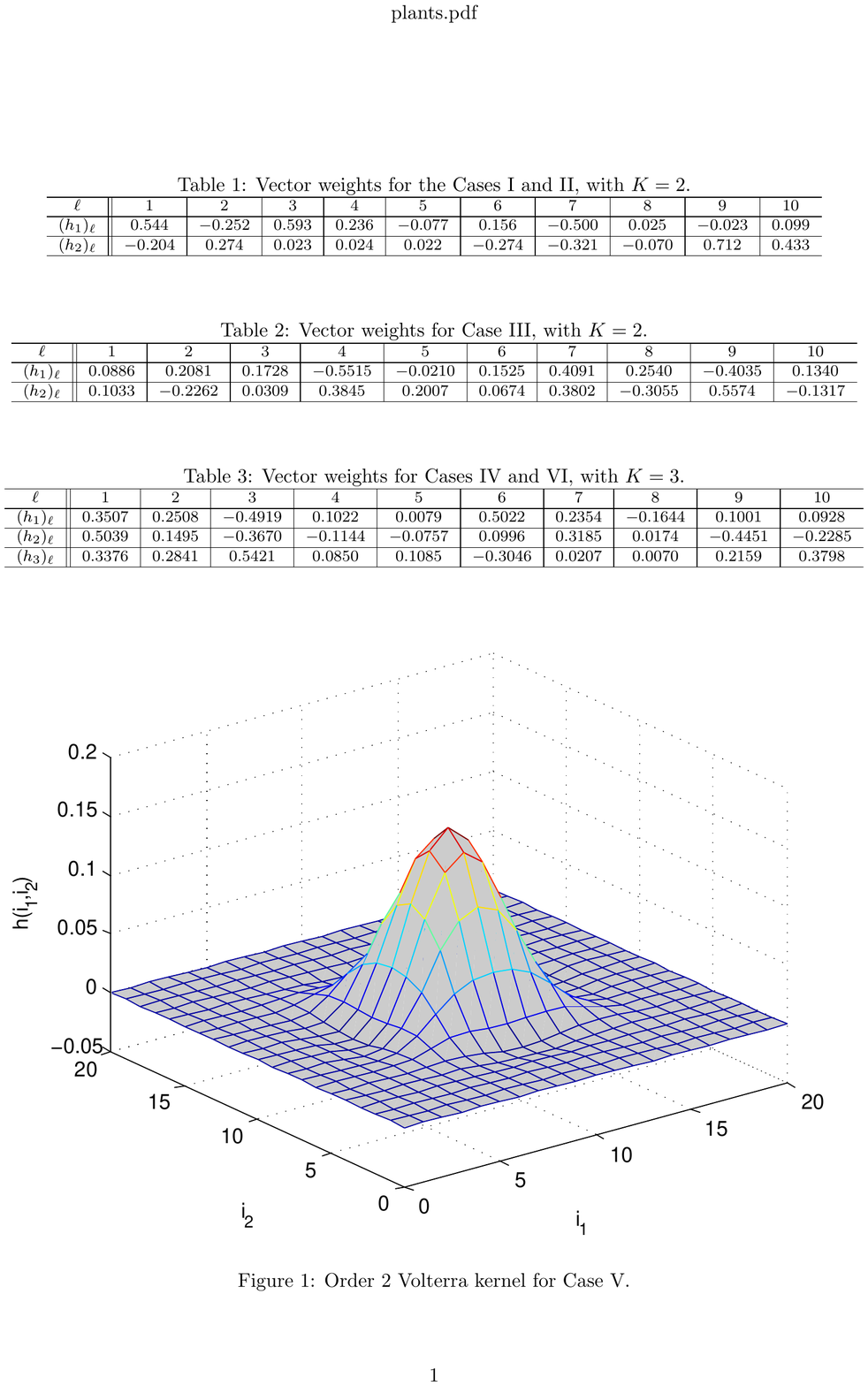}

\end{document}